\documentclass[prl,twocolumn,preprintnumbers]{revtex4}
\usepackage{graphicx}
\begin{document}
%
%
\def\be{\begin{equation}}
\def\ee{\end{equation}}
\def\bc{\begin{center}}
\def\ec{\end{center}}
\def\bea{\begin{eqnarray}}
\def\eea{\end{eqnarray}}
\def\dd{\displaystyle}
\def\nn{\nonumber}
\def\sgn{{\rm sgn}}
\def\ov{\overline}
\preprint{\vbox{\hbox{DFPD-01/TH/31}
\hbox{ROME1-1318-01}}}
\title{Generalized symmetry breaking on orbifolds}
\author{Jonathan A.~Bagger,$^1$
Ferruccio Feruglio$^2$
and
Fabio Zwirner$^3$
}
\affiliation{
$^1$ 
Department of Physics and Astronomy, Johns Hopkins University,
3400 North Charles Street, Baltimore, MD, 21218, USA 
\\
$^2$ 
Dipartimento di Fisica `G.~Galilei', Universit\`a di Padova and
INFN, Sezione di Padova, Via Marzolo~8, I-35131 Padua, Italy
\\
$^3$ 
Dipartimento di Fisica, Universit\`a di Roma `La Sapienza' and
INFN, Sezione di Roma, P.le Aldo Moro~2, I-00185 Rome, Italy}
\date{July 16, 2001}
%
%
\begin{abstract}
We reconsider the phenomenon of mass generation
via coordinate-dependent compactifications
of higher-dimensional theories on orbifolds.  For definiteness,
we study a generic five-dimensional (5D) theory compactified
on $S^1/Z_2$.  We show that the presence of fixed points, 
where the fields or their derivatives may be discontinuous,
permits new realizations of the Scherk-Schwarz mechanism
where, for example, the mass terms are localized at the
orbifold fixed points.  Our technique can be used to
describe the explicit breaking of global flavor symmetries
and supersymmetries by brane-localized mass terms.  It
can also be applied to the spontaneous breaking of local
symmetries, such as gauge symmetries or supergravities.
\end{abstract}
\maketitle
\vspace{0.2in}
%
\section{Introduction}

Coordinate-dependent compactifications of higher-dimensional
theories, first proposed by Scherk and Schwarz \cite{ss}, provide
an elegant and efficient mechanism for mass generation and 
symmetry breaking.   The basic idea is very simple:  one twists
the boundary conditions in the compact extra dimensions by a
global symmetry of the action.  From a four-dimensional (4D) point
of view, this twist induces mass terms that break the symmetries
with which it does not commute.  (For early applications, see
\cite{fayet}.)

In this letter we study Scherk-Schwarz compactifications
of field theories on orbifolds.  We restrict our attention to
compactifications from five to four dimensions on the orbifold
$S^1/Z_2$.  Consistent Scherk-Schwarz compactifications
on this space were first formulated in string theory \cite{fkpz}
and later in field theory \cite{ssft}.  Related phenomenology
was explored in \cite{sspheno}, related field-theoretical
models in \cite{moreft}, and more string realizations 
in \cite{stringss}.

In what follows we present a new type of coordinate-dependent
compactification in which the fields and their derivatives can
jump at the orbifold fixed points.  The discontinuities give
rise to new possibilities for symmetry breaking.  In particular,
they give rise to mass terms that can be localized,
partially or even completely, at the orbifold fixed points.  This
suggests a close connection between our realization and
localized brane dynamics.  Moreover, in contrast to the standard
case where the Scherk-Schwarz mass spectrum is completely
determined by the overall twist, we will find that the
spectrum also depends on the behavior of the fields at the
orbifold fixed points.

Our results have a wide range of applications. They can
be used to generate the explicit breaking of global symmetries,
such as rigid supersymmetry or flavor symmetry. They can
also be used to induce the spontaneous breaking of local
symmetries, such as grand unified gauge symmetries or
supergravity. Indeed, as we discuss in a companion paper
\cite{bfz}, our results encompass such dynamical supersymmetry
breaking mechanisms as gaugino condensation at the
orbifold fixed points \cite{horava}.

The plan of this paper is as follows.  We first explain the general
features of our construction. We then illustrate our results with
a simple example, a free 5D massless fermion
with $U(1)$ twisted boundary conditions.  We conclude with some
comments on the spontaneous breaking of local symmetries, in 
particular supergravity, and on further applications.  

\section{General mechanism}

We consider a generic 5D theory compactified on
the orbifold $S^1/Z_2$, with space-time coordinates $x^M \equiv
(x^m,y)$.  We work on the covering space $S^1$, defined by
identifying the coordinates $y$ and $y + 2  \pi R$, where $R$
is the radius.  We project to the orbifold $S_1/Z_2$ by further
identifying the coordinates $y$ and $-y$.  We denote by
$\Psi(x^m,y)$ all the fields of the 5D theory,
classifying them in representations of the 4D Lorentz group.

We assume that the theory has a continuous global symmetry,
whose action on the fields is given by $\Psi \rightarrow \Psi' =
U \, \Psi$, where $U$ is a unitary matrix.  We define the
$Z_2$ transformations of the fields by
\be
\Psi(-y) = Z  \, \Psi (y) \, ,
\label{z2rep}
\ee
where $Z$ is a matrix such that $Z^2=1$.  It is not restrictive
for us to take a basis in which $Z$ is diagonal,
\be
\label{firep}
\Psi = \left(
\begin{array}{c} 
\Psi^+ \\ \Psi^- 
\end{array}
\right) \, ,
\;\;
Z = diag \, (1,\ldots,1,-1,\ldots,-1) \, .
\label{basis}
\ee

We implement the Scherk-Schwarz mechanism by twisting
the boundary conditions on $S^1$.  Since the fields $\Psi(y)$
are multi-valued on the circle, it is convenient to define the
twist on the real axis:
\be
\label{twist}
\Psi(y)= U_{\vec\beta} \, \Psi(y+2\pi R) \, ,
\ee
where the matrix $U_{\vec\beta}$ depends on the real
parameters ${\vec\beta}$, but not on the space-time
coordinates.  A well-known consistency condition
\cite{fkpz,ssft} between the twist and the orbifold
projection is that
\be
\label{cond2}
U_{\vec\beta} Z U_{\vec\beta} = Z \, .
\ee
If we write $U_{\vec\beta} = \exp (i {\vec\beta}\cdot
\vec T)$, where the matrix ${\vec\beta}\cdot \vec T$ is
hermitian, we see that eq.~(\ref{cond2}) is satisfied if
$\{ {\vec\beta}\cdot \vec T , Z \} = 0$.  This
implies that the generator ${\vec\beta}\cdot \vec T$ is
purely off-diagonal in the basis of eq.~(\ref{basis}).

Our theory is defined on the orbifold $S^1/Z_2$, so we
allow the fields to jump at the orbifold fixed points:
\be
\label{cond11}
\Psi(y_q+\xi)= U_q \, \Psi(y_q-\xi) \, ,
\ee
where $y_q = q \pi R$, $q \in Z$, $0 < \xi \ll 1$ and $U_q$
is a global symmetry  transformation.  The jumps across
points related by a $2\pi R$ translation must be the same,
so
\be
U_{2q} \equiv  U_0 \, ,
\;\;\;
U_{2q+1} \equiv U_\pi \, .
\ee 
A consistency condition identical to (\ref{cond2}) holds
for each of the jumps: 
\be
\label{cond22}
U_q Z U_q = Z \, .
\ee

The physical spectrum is controlled by the Scherk-Schwarz
twist and by the jumps at the orbifold fixed points.  The
discontinuities are the result of mass terms localized
at the fixed points.  In the next section, we shall see that
the mass terms can be described by more than one brane
action.  We will also see that the theory with discontinuities
is equivalent to a conventional Scherk-Schwarz theory with
a modified twist.  In particular, it is possible for the 
discontinuities to completely remove the symmetry breaking 
induced by the twist!

\section{Example}

To illustrate our mechanism in a simple setting, we consider 
the equation of motion for a free 5D massless fermion, written 
in terms of 5D fields with 4D spinor indices
\be
i \sigma^m \partial_m \ov{\Psi} - i \sigma^2 \partial_y \Psi=0 \, ,
\label{eom1}
\ee
valid in each region $y_q<y<y_{q+1}$ of the real axis.
In the notation of eqs.~(\ref{z2rep}) and (\ref{firep}), we
write:
\be
\Psi =
\left(
\begin{array}{c}
\psi_1
\\
\psi_2
\end{array}
\right) \, ,
\;\;\;
\ov{\Psi} =
\left(
\begin{array}{c}
\ov{\psi_1}
\\
\ov{\psi_2}
\end{array}
\right) \, ,
\;\;\;
Z =
\left(
\begin{array}{cc}
1 & 0 
\\
0 & -1
\end{array}
\right) \, .
\ee
The equation of motion (\ref{eom1}) is invariant under global
$SU(2)$ transformations of the form $\Psi'=U \Psi$, where
$U \in SU(2)$.  We take
\be
\label{condspe}
U_\beta = \exp \left( i \beta \sigma^2 \right) 
=
\left(
\begin{array}{cc}
\cos \beta & \sin \beta
\\
- \sin \beta &  \cos \beta
\end{array}
\right) \, ,
\ee
\be
\label{jumpsex}
U_q = \exp \left( i \delta_q \sigma^2 \right) 
=
\left(
\begin{array}{cc}
\cos \delta_q & \sin \delta_q
\\
- \sin \delta_q &  \cos \delta_q
\end{array}
\right) \, ,
\ee
where $\delta_{2q}=\delta_0$ and $\delta_{2q+1}=\delta_\pi$
for any $q \in Z$.

We seek solutions $\Psi(y)$ to eq.~(\ref{eom1}), with the boundary
conditions of eqs.~(\ref{condspe}) and (\ref{jumpsex}). Exploiting
the fact that $i \sigma^m \partial_m \ov{\Psi} = m \Psi$, we find
\be
\label{eigf}
\Psi(y)= \chi \left(
\begin{array}{c}
\cos[m y - \alpha(y)]
\\
\sin[m y - \alpha(y)]
\end{array}
\right) \, ,
\ee
where $\chi$ is a $y$-independent 4D spinor,
\be
\label{spectrum}
m=\frac{n}{R}-\frac{(\beta-\delta_0-\delta_\pi)}{2\pi R}
\, ,
\;\;\;
(n\in Z) \, ,
\ee
and
\be
\alpha(y)=\frac{\delta_0-\delta_\pi}{4}\epsilon(y)
+\frac{\delta_0+\delta_\pi}{4}\eta(y) \, .
\ee
Here  $\epsilon(y)$ is the `sign'  function defined on
$S^1$, and
\be
\label{eta}
\eta(y) = 2 q + 1 \, , 
\;\;
y_q < y < y_{q+1}  \, ,
\;\;
(q\in Z) \, ,
\ee
is the `staircase' function that steps by two units every
$\pi R$ along $y$. The function $\alpha(y)$ satisfies
\be
\label{twistd}
\alpha(y+2\pi R)=\alpha(y)+\delta_0+\delta_\pi\, .
\ee
so the solution (\ref{eigf}) has the correct Scherk-Schwarz
twist.  Sample solutions are shown in Fig.~1.

\begin{figure}
\includegraphics*[width=3in]{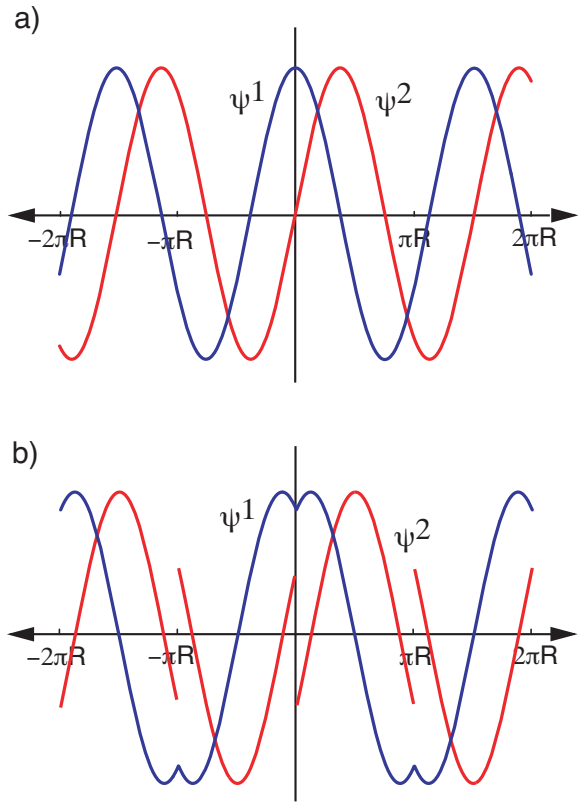}
\caption{Sample profiles for the eigenfunctions $\psi_1$ and
$\psi_2$, with $n=1$, for two cases with identical spectra.  
(a) No jumps:  $\beta=-2$, $\delta_0 = \delta_\pi = 0$.
(b) No twist:  $\beta=0$,
$\delta_0 = \delta_\pi = 1$.}
\label{profile}
\end{figure}

The spectrum (\ref{spectrum}) is characterized by a uniform
shift with respect to a traditional Kaluza-Klein compactification.
In contrast to the usual Scherk-Schwarz mechanism, however,
the shift depends on the jumps $\delta_0$ and $\delta_\pi$,
as well as on the twist $\beta$.   In particular, it is possible to
have a vanishing shift for nonvanishing $\beta$.  In the limit
$\delta_q\to 0$, our results reduce to the conventional
Scherk-Schwarz spectrum.  Note that the eigenfunction of
eq.~(\ref{eigf}) is discontinuous:  the even part has cusps
and the odd part has jumps at $y=y_q$, as required by the
boundary conditions.  In the limit $\delta_q \to 0$ the
eigenfunction becomes regular everywhere. 

For any $\delta_q$, the system is equivalent to a conventional
Scherk-Schwarz compactification with twist  $\beta_c = \beta
-\delta_0-\delta_\pi$.  The new field variable, $\Psi_c$, is related
to the discontinuous variable, $\Psi$, via the generalized function 
$\alpha(y)$,
\be
\label{redef}
\left(
\begin{array}{c}
\psi_{1c}\\
\psi_{2c}
\end{array}
\right) = 
\left(
\begin{array}{cc}
\cos \alpha(y) & \sin \alpha(y)
\\
- \sin \alpha(y) &  \cos \alpha(y)
\end{array}
\right) \, 
\left(
\begin{array}{c}
\psi_{1}\\
\psi_{2}
\end{array}
\right) \ .
\ee
This is reminiscent of strong CP violation, where the physical
order parameter is not $\theta$, but the combination $\theta - 
{\rm arg}\, {\rm det}\, m_q$, where $m_q$ is the quark mass
matrix.  Similarly, the mass shift of our system is controlled not
by $\beta$ alone, but by the twist $\beta_c$, which includes
contributions from jumps in the fermion fields.  As in QCD,
where we can eliminate the phase in ${\rm det}\, m_q$ by
a chiral transformation, here we can remove the jumps by
a redefinition of the fermion fields.  In the new basis,
there are no jumps, but the twist acquires an additional
contribution.

Discontinuous field variables arise from mass terms localized
at the fixed points.  This can be seen by starting with the 
Lagrangian ${\cal L}$ for the fermions $\psi_c^{1,2}(y)$,
characterized by a twist $\beta_c=\beta-\delta_0-\delta_\pi$:
\bea
{\cal L}(\psi_c)&=&
 i \ov{\psi_c^1} \ov{\sigma}^m \partial_m \psi_c^1 
+ i \ov{\psi_c^2} \ov{\sigma}^m \partial_m \psi_c^2 
\nn \\
& + &
 \left[ {1 \over 2} \left( \psi_c^2 \partial_y \psi_c^1 
- \psi_c^1 \partial_y \psi_c^2 \right) + {\rm h.c.} \right]  \, .
\label{lagr}
\eea
If we perform the field redefinition of eq.~(\ref{redef}),
the 5D Lagrangian becomes:
\be
{\cal L}(\psi_c)={\cal L}(\psi)+{\cal L}_{brane}(\psi)\, ,
\label{ldecu}
\ee
where
\bea
\label{lbr1}
{\cal L}_{brane}(\psi)&=&-\frac{1}{2} \alpha'(y) 
\left( \psi_1 \psi_1 + \psi_2 \psi_2 \right)
+ {\rm h.c.}\, ,
\eea
and
\be
\alpha'(y)=\sum_{q=-\infty}^{+\infty}
\left[\delta_0 \, \delta(y- y_{2q}) 
+ \delta_\pi \, \delta(y- y_{2 q+1})\right]
\, .
\label{dprime}
\ee
We see that the jumps $\delta_q$ arise from fermion mass terms
localized at the orbifold fixed points.

The discontinuities of the fields can be recovered by integrating
the equations of motion. The trick is to find the correct equations.
We avoid all subtleties associated with discontinuous field
variables by {\it defining} the term that appears in the brane
action to be {\it continuous} across the orbifold fixed points. 
For the case at hand, this means one must choose the field
variables so that the combination $\psi_1\psi_1+ \psi_2\psi_2$
is continuous.  Alternatively, one can obtain the equations of
motion by first regularizing the delta functions, so that  $\psi_1$
and $\psi_2$ are continuous, and then taking the singular
limit.

It is interesting to note that the same physical system can
be obtained from another brane Lagrangian, one in which
we treat the even field $\psi_1(y)$ as continuous.   The
discontinuity of the odd field $\psi_2(y)$ is then
\be
\psi_2(y_q+\xi)-\psi_2(y_q-\xi)=-2 \tan\frac{\delta_q}{2}\,
\psi_1(y_q) \, .
\ee
This jump is reproduced by the brane Lagrangian
\be
\label{lbr2}
{\cal L}\,'_{brane}(\psi)=-\frac{1}{2} f(y)  \,
\psi_1 \psi_1 + {\rm h.c.} \, , \ee where \be f(y)=2 \sum_{q \in Z}
\left[\tan\frac{\delta_0}{2} \, \delta(y \! - \! y_{2q}) +
\tan\frac{\delta_\pi}{2} \, \delta(y \! - \! y_{2 q+1})\right] \, .
\ee 
In this case, we vary with respect to $\psi_1(y)$ and
$\psi_2(y)$; the discontinuous field $\psi_2(y)$ does not
appear in the brane Lagrangian.

In summary, the brane Lagrangians (\ref{lbr1}) and (\ref{lbr2})
give rise to equivalent theories in the absence of brane
interactions, provided we use an appropriate procedure
to derive the equations of motion.

\section{Conclusions}

In this letter we have studied coordinate dependent
compactifications of field theories on orbifolds.  We have seen
that the mass spectrum depends on an overall twist of the fields,
together with the jumps of the fields at the orbifold fixed points.
Such compactifications can break the symmetries of a theory,
either global and local.  The order parameter is nonlocal, in the
sense that it is determined by a combination of the twist and the
discontinuities.  

In a supersymmetric Yang-Mills theory, for example, the twist
and the jumps are defined by a $U(1)_R$ subgroup of $SU(2)_R$.
{}From a 4D point of view, this typically breaks the $N=1$
supersymmetry that survives the orbifold projection.  Note,
though, that it is possible for supersymmetry to remain unbroken.
For instance, when $\beta=0$, supersymmetry is preserved in
the presence of opposite, nonvanishing jumps at $y=y_{2q}$
and $y=y_{2q+1}$, in analogy with a phenomenon first discussed
in $M$-theory \cite{horava}.
This example can be readily extended to the case where the
fermions $\psi_1$ and $\psi_2$ come in $n$ distinct copies, in
which case flavor symmetry is broken if the matrices $U_{\beta}$
and $U_q$ have a non-trivial structure in flavor space. 

It is important to note that our mechanism provides a self-consistent
way of introducing other interaction terms, such as Yukawa couplings,
or even kinetic terms, that are localized at the fixed points. Such 
terms will always occur in non-renormalizable theories, including 
supergravity, where the kinetic terms typically have a non-canonical 
(and non-renormalizable) form.

It would be interesting to find string realizations of our
mechanism, which so far are missing. These would give rise
to models where mass terms for the untwisted fields are
localized at the fixed points of a non-freely acting orbifold.

\section{Acknowledgements}
We thank D.~Belyaev, K.~Dienes, A.~Hebecker, J.-P.~Hurni, E.~Kiritsis, C.~Kounnas, 
A.~Masiero, M.~Porrati, L.~Silvestrini, C.~Scrucca, M.~Serone and 
N.~Weiner for discussions. We especially thank C.~Biggio for her 
valuable help in improving the first version of the manuscript. We 
also thank the Aspen Center of Physics, where part of this work was 
done, for its warm hospitality.  J.B.\ is supported by the U.S. National 
Science Foundation, grant NSF-PHY-9970781. F.F.\ and\ F.Z. are partially
supported by the European Program HPRN-CT-2000-00148.
\vspace*{0.5cm}
\end{document}